\documentclass[12pt]{article}

\usepackage{scicite}

\usepackage{times}

\usepackage{graphicx}

\newenvironment{sciabstract}{%
\begin{quote} \bf}
{\end{quote}}

\newcounter{lastnote}
\newenvironment{scilastnote}{%
\setcounter{lastnote}{\value{enumiv}}%
\addtocounter{lastnote}{+1}%
\begin{list}%
{\arabic{lastnote}.}
{\setlength{\leftmargin}{.22in}}
{\setlength{\labelsep}{.5em}}}
{\end{list}}

\title{Supernova Shock Breakout from a Red Supergiant}

\author {Kevin Schawinski,$^{1\ast}$ 
Stephen Justham,$^{1\ast}$
Christian Wolf,$^{1\ast}$ \\ 
Philipp Podsiadlowski,$^{1}$ 
Mark Sullivan,$^{1}$ 
Katrien C.~Steenbrugge,$^{2}$ \\ 
Tony Bell,$^{1}$ 
Hermann-Josef R\"{o}ser,$^{3}$ 
Emma S.~Walker,$^{1}$  
Pierre Astier,$^{4}$\\
Dave Balam,$^{5}$
Christophe Balland,$^{4}$
Ray Carlberg,$^{6}$
Alex Conley,$^{6}$\\
Dominque Fouchez,$^{7}$
Julien Guy,$^{4}$
Delphine Hardin,$^{4}$
Isobel Hook,$^{1}$\\
D.~Andrew Howell,$^{6}$
Reynald Pain,$^{4}$
Kathy Perrett,$^{6}$
Chris Pritchet,$^{5}$\\
Nicolas Regnault$^{4}$
and Sukyoung K.~Yi$^{8}$\\
\\ \normalsize{$^{1}$Department of Physics, University of Oxford,
Oxford OX1 3RH, UK.}\\
\normalsize{$^{2}$St John's College Research
Centre, University of Oxford, OX1 3JP, UK. }\\
\normalsize{$^{3}$Max-Planck-Institut f\"{u}r Astronomie,
K\"{o}nigstuhl 17, 69117 Heidelberg, Germany.}\\
\normalsize{$^{4}$ LPHNE, CNRS-IN2P3 and Universit\'{e}s Paris VI \& VII, 4
Place Jussieu,}\\
\normalsize{75252 Paris Cedex 05, France.}\\
\normalsize{$^{5}$ Department of Physics and Astronomy, University of
Victoria, }\\
\normalsize{ PO Box 3055 STN CSC, Victoria, BC V8T 3P6, Canada.}\\
\normalsize{$^{6}$ Department of Physics and Astronomy, University of
Toronto,}\\ 
\normalsize{50 St. George Street, Toronto, ON M5S 3H4, Canada.}\\
\normalsize{$^{7}$ CPPM, CNRS-IN2P3 and Université Aix-Marseille II,
Case 907,}\\
\normalsize{13288 Marseille Cedex 9, France.}\\
\normalsize{$^{8}$Department of Astronomy, Yonsei University, Seoul
120-749, Korea.}\\ \\
\normalsize{$^\ast$To whom correspondence should be addressed;} \\
\normalsize{E-mail: kevins@astro.ox.ac.uk, sjustham@astro.ox.ac.uk,
cwolf@astro.ox.ac.uk} }

\date{}

\begin{document}


\baselineskip24pt

% Macros
\newcommand\aj{{AJ}}
\newcommand\araa{{ARA\&A}}
\newcommand\apj{{ApJ}}
\newcommand\apjl{{ApJ}}
\newcommand\apjs{{ApJS}}
\newcommand\aap{{A\&A}}
\newcommand\nat{{Nature}}
\newcommand\mnras{{MNRAS}}
\newcommand\pasp{{PASP}}
\newcommand\physrep{{Phys. Rep.}}


\maketitle


\begin{sciabstract}
Massive stars undergo a violent death when the supply of nuclear fuel
in their cores is exhausted, resulting in a catastrophic
"core-collapse" supernova. Such events are usually only detected at
least a few days after the star has exploded. Observations of the
supernova SNLS-04D2dc with the Galaxy Evolution Explorer space
telescope reveal a radiative precursor from the supernova shock before
the shock reached the surface of the star and show the initial
expansion of the star at the beginning of the explosion. Theoretical
models of the ultraviolet light curve confirm that the progenitor was
a red supergiant, as expected for this type of supernova. These
observations provide a way to probe the physics of core-collapse
supernovae and the internal structures of their progenitor stars.
\end{sciabstract}

The explosive deaths of massive stars are dramatic events that seed
the Universe with heavy elements (\textit{1, 2}), produce black holes,
pulsars, and the most energetic gamma-ray bursts (GRBs; \textit{3}).
Their energy input can regulate the growth of galaxies
(\textit{4}). Even though a large amount of theoretical effort has
been expended on trying to explain how the terminal collapse of a
star's core leads to a luminous supernova, we do not fully understand
the process by which the collapse of the core produces an
outward-moving shock that leads to the ejection of the envelope
(\textit{5--7}). This shock heats and accelerates the stellar envelope
as it passes through it. By the time the shock dissipates at the
surface of the star, several solar masses of previously static
envelope material are expanding at a few percent of the speed of
light. At the time of core collapse, a nearby external observer
equipped with a detector of neutrinos or of gravitational waves might
receive a brief warning of the future explosion, but for most of the
passage of the shock through the star that observer would notice no
further change. Only when the shock approaches the surface does
radiation diffuse far enough ahead of the shock wave to raise the
temperature of the stellar photosphere.  This phase is sometimes
referred to as `shock breakout', although the associated radiation is
from the `radiative precursor' of the shock, long before the shock
actually reaches the surface. This radiative precursor raises the
temperature of the star to $\sim 10^{5}$ K before the surface expands
dramatically (\textit{8}).

Shock breakouts have been inferred for a few relatively local GRBs and
x-ray flashes which may involve shocks traveling through dense winds
outside compact blue stars, including the recent SN 2008D
(\textit{9--14}). Here we describe the brightening of a red supergiant
due to the theoretically predicted radiative precursor before the
supernova shock reaches the surface of the star. Such observations
provide information about the density profile inside the progenitor
star (\textit{15}) and the physics of radiative shocks, and knowledge
of the spectrum of the associated ultraviolet flash has implications
for the ioniziation of the circumstellar medium (\textit{16, 17}).

Although core-collapse supernovae are expected to be most luminous
around the time of shock breakout, most of this energy emerges as
extreme UV or soft X-ray radiation. Hence core-collapse supernovae are
typically only discovered several days after the supernova explosion
near the peak of their optical light curve; observations of early
light curves are rare (\textit{18, 19}). To circumvent this problem,
we exploit two complementary data sets: an optical survey to locate
supernovoae, and UV data to search for serendipitously associated
shock breakouts. The first is the Supernova Legacy Survey (SNLS;
\textit{20}) which studies distant supernovae using data taken every 4
days at the 3.6~m Canada-France-Hawaii Telescope (CFHT). The second is
from the Galaxy Evolution Explorer (\textit{GALEX}) UV space telescope
(\textit{21, 22}), which took a deep 100 hour combined exposure
coincident with the early-2004 SNLS data in the Cosmological Evolution
Survey I``COSMOS'') field (\textit{23, 24}).  The \textit{GALEX} data
were taken using sub-exposures of 15 to 30 minutes over several weeks,
providing data with the time resolution necessary to resolve
UV-luminous events occurring before the SNLS supernovae.

One SNLS event, designated SNLS-04D2dc and confirmed as a Type II
supernova from the hydrogen lines in an optical spectrum taken at the
European Southern Observatory (ESO) Very Large Telescope (VLT)
[supporting online material S1 (SOM text S1) (\textit{25})], shows a
dramatic brightening in the \textit{GALEX} near-UV images about 2
weeks before the discovery by the SNLS, consistent with shock
breakout. The host galaxy appears to be a normal star-forming spiral
galaxy at a redshift of $z = 0.1854$. The supernova spectrum, Gemini
host galaxy spectrum and \textit{Hubble Space Telescope} image of the
host are presented in \textit{25}. The optical light curve has a
plateau that identifies the explosion as a Type IIP supernova
(Fig.~1), suggesting a red-supergiant progenitor (\textit{26,
27}). Because of bad weather and technical problems with the CFHT
camera, there are no optical data concurrent with the UV data;
however, \textit{GALEX} observed the entire radiative precursor,
(Fig.~2).

The \textit{GALEX} light curve probes the arrival of the supernova
shock at the surface of the star.  We can interpret the two peaks in
this light curve (Fig.~2) in terms of distinct physical processes. The
first peak in the UV light curve is due to radiation traveling ahead
of the shock wave. This heats the surface of the star before it begins
to explode. The near-UV light curve samples the brightening caused by
this precursor over 6 hours. We can compare the duration of the
observed precursor with theoretical expectations by equating the
photon diffusion time-scale with the time-scale for the shock to
escape from the envelope. If $v$ is the shock speed and the density of
the hydrogen-dominated atmosphere is $\rho$, we find $d \approx 2.5
\times 10^{11}~{\rm{m}} (10^{-8}~{\rm kg~m^{-3}}/\rho) (10^{7}~{\rm
m~s^{-1}}/v)$ for the depth of the shock $d$ (from the surface of the
star) at the time when the radiative precursor becomes visible at the
surface (SOM text 3 and 4). This value for $d$ leads to a prediction
for the duration of the shock precursor of $d/v = 2.5 \times
10^{4}~{\rm s}$ for the parameters above, that is, almost 7 hours,
consistent with our observations of the precursor. This indicates that
the progenitor was a large star, that is, a red supergiant, as
expected for the progenitor of a type IIP supernova (\textit{26, 27}),
whilst previous calculations indicated that radiative precursors from
blue supergiant stars would last for minutes rather than hours
(\textit{8}). To model the radiative precursor, we solved simplified
radiation-hydrodynamics equations for an outward-moving shock inside a
stellar envelope. Figure 3 shows representative models that are
consistent with the data; they require radii and envelope densities
appropriate for a red-supergiant star. These models also indicate that
only the initial $\sim 4$ hours of the first UV peak occur before the
shock reaches the surface of the star.

The peak in the total luminosity of the source occurs at the time of
the first UV peak, and the total luminosity monotonically decreases
after this point. The temperature behind the shock is lower than the
temperature at the shock front itself, which leads to a rapid drop in
the luminosity of the star after shock breakout (\textit{8}). The
near-UV light curve in Fig.~3 shows this dip in brightness after the
shock has escaped from the star.

Although the radiative precursor does cause some expansion of the
star, there is little change in the stellar radius until the shock
reaches the surface. Behind the shock, the radiation-dominated plasma
expands at almost constant velocity and cools rapidly as a result of
adiabatic expansion {\em (1)\/}. The UV light curve is now governed by
the expansion of the photospheric radius (and concomitant increase in
radiating surface area), the adiabatic cooling of the surface and the
shift of the spectral energy distribution towards longer wavelengths,
causing the second peak in the UV light curve. In the adiabatic
cooling phase, the photospheric temperature $T$ is approximately
inversely proportional to the photospheric radius $R$. Because for a
black body this drop in $T$ causes a more rapid decrease in the
luminosity ($L\propto T^4\propto R^{-4}$) than the increase due to the
growing surface area ($L\propto R^2$), the total luminosity of the
supernova continues to decrease. However, in the Rayleigh-Jeans
portion of the spectrum the increase in the surface area of the
photosphere is more important than the decrease in emission per unit
area, and the luminosity at those wavelengths increases (SOM text
S4.3). The observed UV luminosity rises until the peak of the
blackbody spectral energy distribution nears the UV
waveband. Thereafter, the UV luminosity decreases with continued
adiabatic expansion and cooling. The model curves in Fig.~2 show that
this simple physical description reproduces the \textit{GALEX} data
with parameters as expected for a red supergiant progenitor. Initial
photospheric radii of 500-- to 1000 solar radii $R_{\odot}$, expansion
velocities of 1-- to2$\times 10^{7}~{\rm m~s^{-1}}$, and initial
temperatures of $\sim 10^{5}~{\rm K}$ match the observed fluxes
well. The biggest uncertainty arises from the adopted extinction (SOM
text S1); any increase in the NUV extinction would increase the range
of preferred initial radii. Measuring precise radii of supernova
progenitor stars would be a valuable constraint of the late stages in
the evolution of massive stars; this require higher time resolution
and more accurate temperature determinations, for example, from
observing the full spectral energy distribution from x-ray to
optical. In addition, detailed light curves of radiative precursors
probe the energetics of supernova shocks and the structures of the
stellar envelopes through which they travel.

% Following is a new environment, {scilastnote}, that's defined in the
% preamble and that allows authors to add a reference at the end of the
% list that's not signaled in the text; such references are used in
% *Science* for acknowledgments of funding, help, etc.

\begin{scilastnote}
\item KS is supported by the Henry Skynner Junior Research Fellowship
  at Balliol College, Oxford. SJ acknowledges support by STFC \&
  Global Jet Watch, CW and ESW by STFC and MS by the Royal
  Society. This work is supported by Acceleration and Basic research
  programs of MOST/KOSEF to SKY. We gratefully acknowledge use of data
  from the NASA \textit{GALEX (Galaxy Evolution Explorer)} satellite,
  the Canada-France-Hawaii Telescope (CFHT), the ESO Very Large
  Telescopes (VLT), the Gemini Observatory and the \textit{Hubble
  Space Telescope}.
\end{scilastnote}

\clearpage
\begin{center}
\includegraphics[angle=90, width=0.95\textwidth]{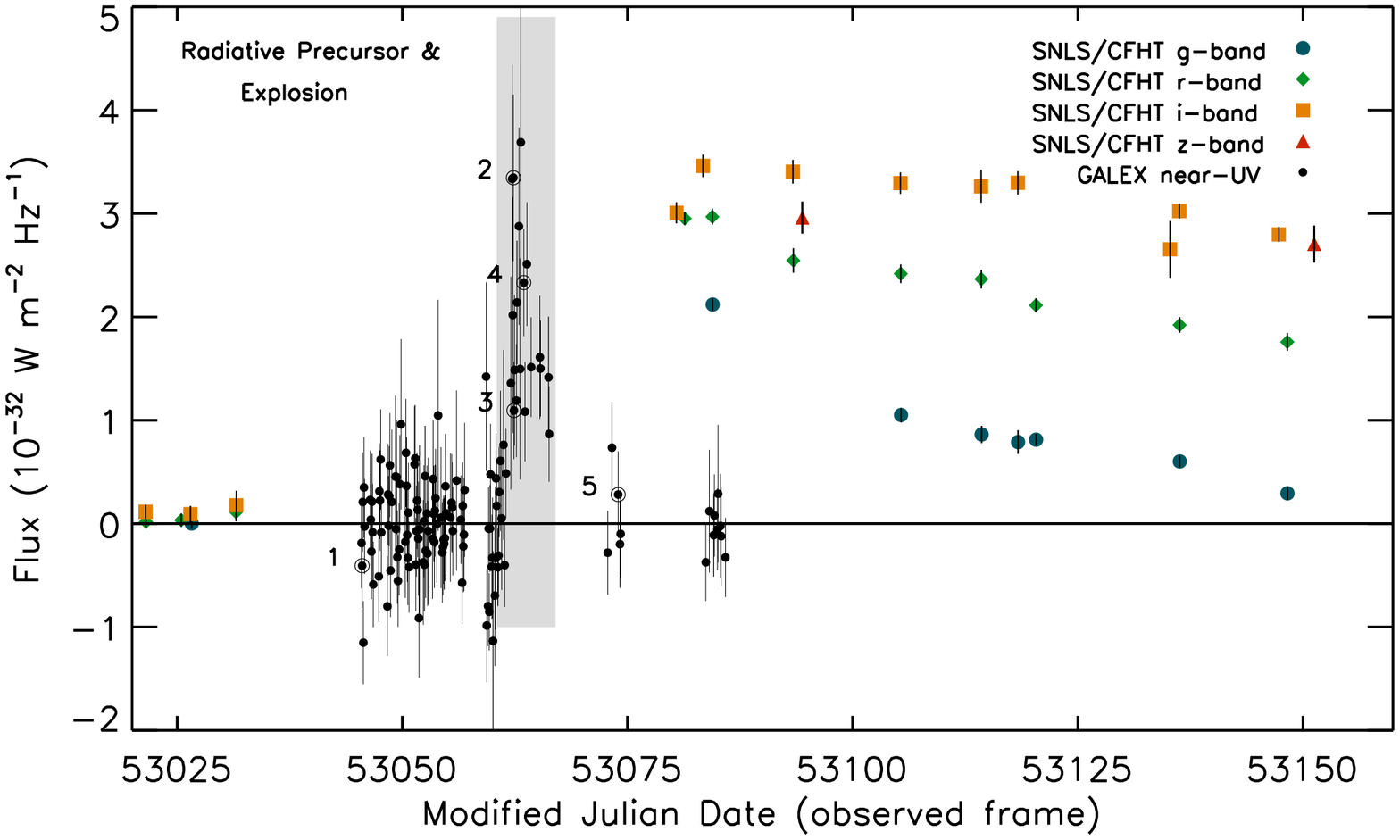}\\
\includegraphics[angle=90, width=0.75\textwidth]{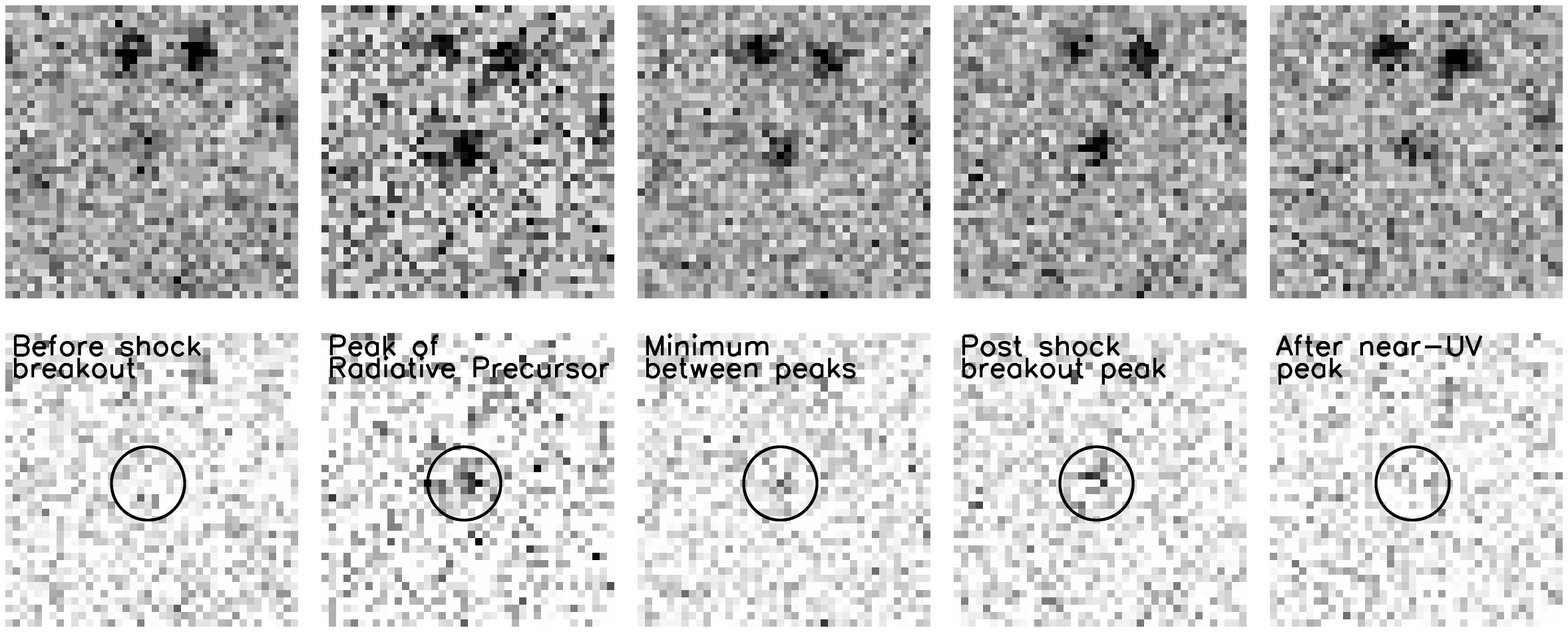}\\
\end{center}
\noindent {\bf Fig.~1.} Composite of the optical SNLS and the UV
\textit{GALEX} light curves, or observed fluxes as a function of
modified Julian date. All fluxes are host galaxy subtracted and are
not corrected for internal extinction. The gray box indicates the time
of the radiative precursor. The points highlighted by circles indicate
five phases of the radiative precursor in the UV, as observed by
\textit{GALEX}. These five phases are illustrated by a time sequence
of original near-UV images (upper row, 1 $\times$ 1 arcmin) and
difference images (lower row, with a pre-SN image subtracted) to
emphasize the transient source. Note the drop from point 2 to the
minimum at 3 and the rise to 4, clearly visible in the \textit{GALEX}
images. The lack of optical data during the UV event is due to both
poor weather conditions and technical problems. Both \textit{GALEX}
and SNLS light curves are available as Tables in the SOM.

\clearpage
\begin{center}
\includegraphics[angle=270, width=\textwidth]{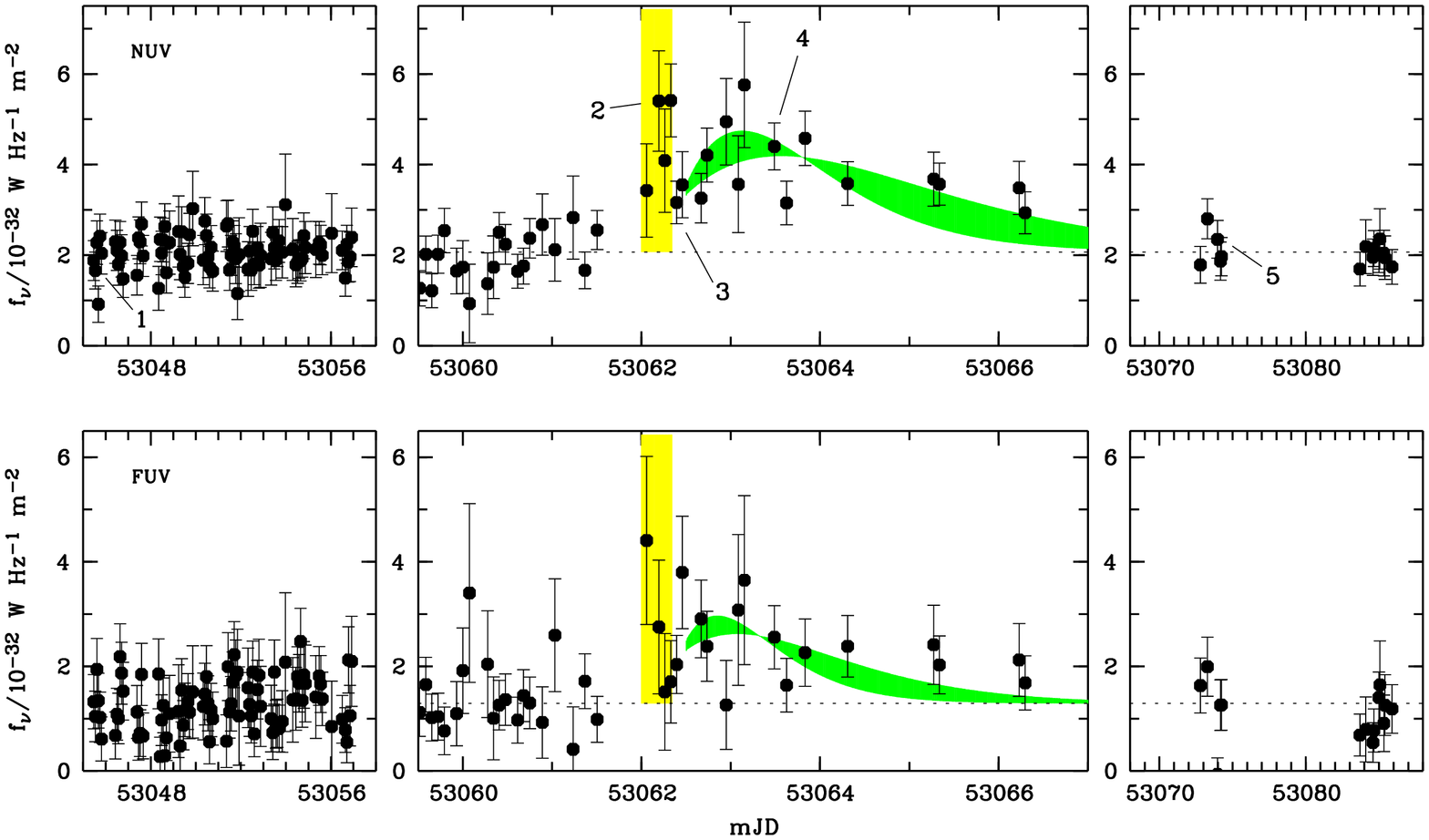}\\
\end{center}
\noindent {\bf Fig.~2.} The \textit{GALEX} near-UV and far-UV flux
against time (modified Julian date in days). This is a zoomed-in
version of the shaded time range of Fig.~1 and we mark the same five
data points. The background levels are shown before and after the
supernova (left and right panels), and the central panels show the
event itself. The radiative precursor is highlighted in yellow.
Models for the post-explosion expansion are shaded in green; these
models assume an initial photospheric radius of 500--1000
$R_{\odot}$. The width of the green band is due to the range of
assumed expansion velocities ($1-2 \times 10^{7}~{\rm
m~s^{-1}}$). These models assume adiabatic free expansion of a
radiation-dominated plasma and black-body emission from a well-defined
photosphere (see text). The models were only fitted to the NUV data,
but are also consistent with the FUV data.

\clearpage
\begin{center}
\includegraphics[width=\textwidth]{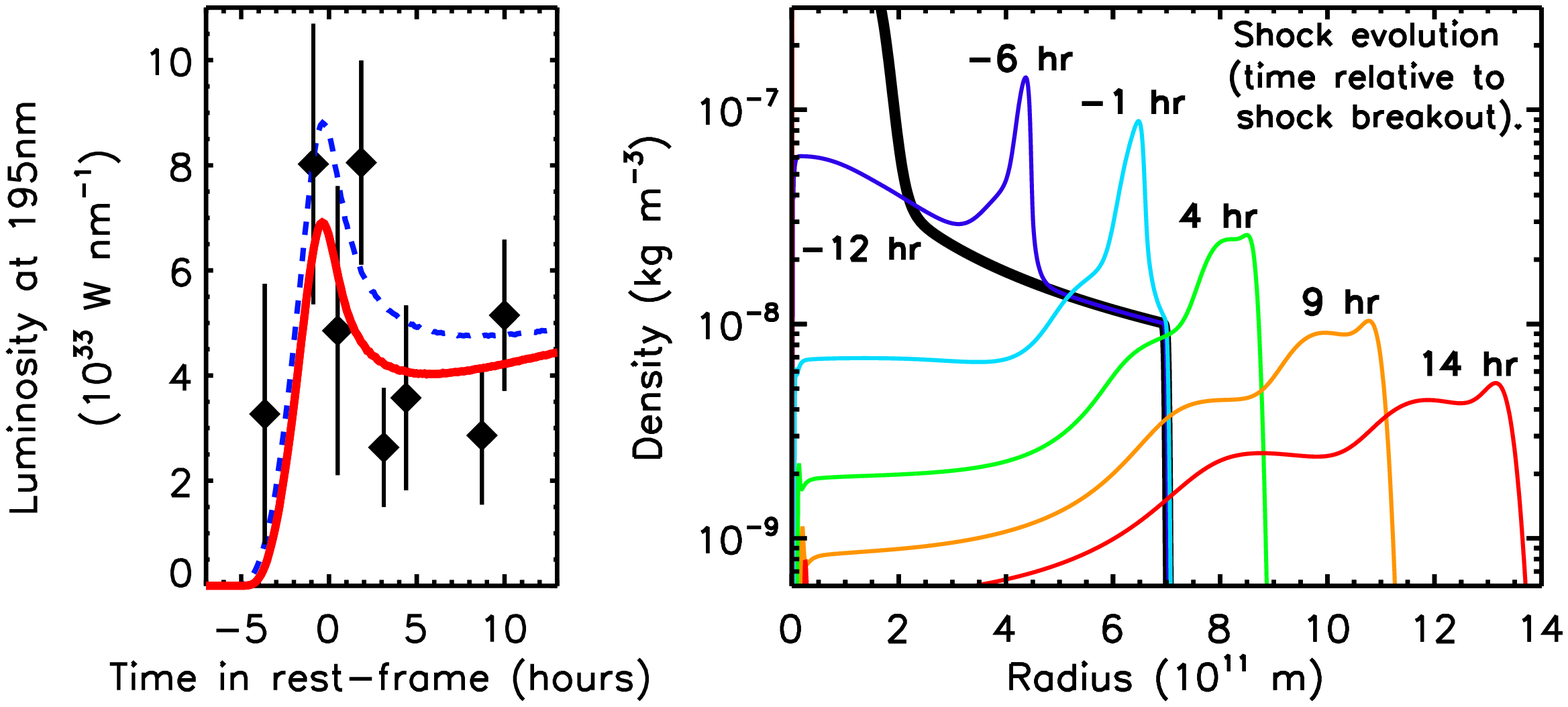}\\
\end{center}
\noindent {\bf Fig.~3.} \textbf{Left:} representative model precursor
light curves compared to the precursor data (diamonds with error
bars). The solid red curve represents the light curve produced from a
model with an initial radius of $7 \times 10^{11}~{\rm{m}}$ ($\approx
1000 R_{\odot}$) and an initial density distribution $\propto
1/r$. The dashed curve in the left hand panel shows an identical model
except that the initial radius was $10^{12}~{\rm{m}}$; the absolute
normalisation of these luminosities is uncertain to factors of order
unity (see supplementary material for details). The zero-point of the
time axis is approximately at shock breakout. \textbf{Right:} the time
evolution of the internal density profile for the model producing the
\emph{solid} light curve in the left panel. Around 4 hours before the
shock has reached the surface the UV luminosity has already begun to
rise.

%%%%%%%%%%%%%%%%%%%%%%%%%%%%%%%%%%%%%%%%%%%%%%%%%%%%%%%%%%%%%%%%%%%%%%%%%%%

\pagebreak

\noindent{\LARGE{Supplementary Online Material}}\\

\section*{S1 Spectra of Supernova and Host Galaxy}

SNLS searches for high-redshift supernovae (SNe) with the goal of
measuring about 500 distant Type Ia supernovae (SNe Ia) for use in
studying dark energy. The volume and depth of these observations also
allows detection of a large number of core-collapse SNe. While the
dedicated spectroscopic follow-up of SNLS is prioritised for the SN Ia
candidates, a spectrum for SNLS-04D2dc (located at RA = 10:00:16.7,
Dec = +02:12:18.52, J2000) was obtained on 2004-03-19 (a few days
after discovery) at the European Southern Observatory Very Large
Telescope.\footnote{ESO/VLT large programme number 171.A-0486.} The
spectrum provided an excellent match to a type II SN (Fig. S1) located
in a star-forming galaxy at a redshift $z=0.1854$. A second spectrum
obtained in early 2008 at the Gillett Gemini North Telescope after the
SN had faded confirms the presence of strong emission lines in the
spiral host galaxy. In Figs.  S1 and S2, respectively, we present the
spectra and the \textit{Hubble Space Telescope} F814W image of the
host galaxy from the COSMOS survey.

We have extracted a spectrum at the location of the SN inside the
galaxy to estimate the host galaxy extincton based on the Balmer
decrement of the galaxy spectrum. We used the SMC reddening law
(\textit{S1}) to estimate the extinction in the GALEX bands, which is
appropriate for most star-forming galaxies at most redshifts including
lower-mass galaxies at low redshift. The colour excess $E_{B-V}$ was
determined from the Balmer decrement at the location of the SN and
gives $E_{B-V} = 0.14$, with implied extinction on the GALEX UV
filters of $A_{\rm NUV}=1.45$\,mag and $A_{\rm FUV}=2.39$\,mag. The
largest source of uncertainty in the extinction measurement is
probably its low spatial resolution. From the variation in $E_{B-V}$
we see between stars within $\sim 1{\rm~kpc}$ of the Sun, we estimate
that the uncertainty in the absolute extinction of the supernova could
be as much as a factor of two.

An alternative way to estimate the extinction of this supernova is to
use the empirical relation found by (\textit{S2}) for a set of IIP
supernovae, based on the colours during the optical plateau. Using the
restframe V-I colour at day 50 of 0.54, their relation is broadly
consistent with the level of extinction for this supernova which we
measure from the emission lines, though their relation is also
consistent with no extinction for this supernova.

\section*{S2 \textit{GALEX} Data Reduction}

We determined the photometry for the NUV light curve by processing
image frames of $10' \times 10'$ size centered on the SN using the
MPIAPHOT package (\textit{S3}). We co-added a selection of 161 frames
with a reasonably Gaussian PSF to obtain best-possible position
estimates when searching objects with SExtractor (\textit{S4}). We
then transformed the coordinates of the object list back into the
coordinate frames of each single exposure and measured fluxes centered
on the projected object positions.

We suppressed the propagation of variations in the PSF (presumably
mostly due to focus drifts) into the photometry by making sure that we
always probe the same physical footprint $f(x,y)$ of any object in all
exposures irrespective of the PSF $p(x,y)$. Here, the footprint
$f(x,y)$ is the convolution of the PSF $p(x,y)$ with the aperture
weighting function $a(x,y)$. If all three are Gaussians, an identical
physical footprint can be probed even when the PSF changes, simply by
adjusting the weighting function $a(x,y)$ for each frame. We chose to
measure fluxes on a footprint of $7.5"$ FWHM, so that on average
$p\approx a$, which optimizes the signal-to-noise ratio of point
sources.

Hence, we measure the PSF on each individual frame, choose the
weighting function needed to conserve the footprint and obtain the
flux on the footprint. Individual frames are normalized to each other
using the count rates of the 15 brightest non-variable objects. Fluxes
from individual frames are averaged for each object and the flux error
is derived from the scatter. Thus, it takes not only photon noise into
account, but also sub-optimal flat-fielding, errors in the background
determination and uncorrected detector artifacts. All fluxes are
finally calibrated to the GALEX photometric catalogues using the
brightest stars. As such the MPIAPHOT aperture fluxes correspond to
total fluxes for point sources, but underestimate them for extended
sources. In this way, we have measured the NUV flux from the SN alone
(the excess flux over the host level) in a physically non-variable
aperture, with ideal S/N, and correctly calibrated.

Fig.~S3 shows the distribution of NUV object flux
scatter vs. mean fluxes for $\sim 1000$ objects detected in the
selected area.  Only two objects, the SN and a QSO, appear as
significantly variable by showing more flux scatter among the frames
than expected from Poissonian noise. When sources are fainter than the
background, their flux scatter is driven by constant noise in the
estimated background (horizontal arm). The scatter of bright sources
is dominated by Poisson fluctuations in the source flux, so these form
a steep arm at slope 1/2 (in a log-log plot).

Fig.~2 shows part of the resulting light curve; a large group of
frames obtained two years after the SN event is omitted as it shows
just the host galaxy light at the same level and scatter as before the
event. The dip between peaks is statistically significant; by summing
the NUV making up the first peak, the dip and the start of the second
peak, we determine that the significance of the drop is 2.77$\sigma$. 

The FUV images have been processed in a similar fashion except that
the low count rates are a challenge for determining the PSF,
background and normalization of individual frames. The resulting FUV
fluxes should be considered uncertain at a +/- 30\% level in each
frame. For this reason the FUV light curve is only shown to indicate
that excess fluxes are observed exactly at the time of the NUV event,
but the FUV data are not included in the model fit for the
photospheric expansion. The model curves are plotted in the FUV panels
just as they are predicted by the NUV fit.

Given that we find two variable sources in an area of 10' x 10' the
probability for a chance coincidence of a random variable with the
location of the host galaxy (known to 1" x 1") is 2:360,000. However,
if you restricted yourself to the short time period near the
supernova, this probability would shrink even further.

We use a cosmology consistent with the WMAP 3 year results and assume
a Hubble Constant $H_{0} = 70$.

\section*{S3 Analytic Estimates for the Radiative Precursor}

Here we give more details about the derivation of the scaling
relations given in the main text.

\subsection*{S3.1 Estimated Depth and Duration}

First we estimate the depth of the shock $d$ within the star at the
start of the radiative precursor. This is defined by equating the
photon diffusion time-scale $\tau_{\rm diff}$ with the time-scale for
the shock to escape $\tau_{\rm s}$, where
\begin{equation}
\tau_{\rm diff} \approx \frac{3 d^{2}}{\alpha~l~c} ,
\end{equation}
\begin{equation}
\tau_{\rm s} \approx \frac{d}{v} , 
\end{equation}
and $l$ is the photon mean free path, $v$ is the shock speed, $c$ is
the speed of light and $\alpha$ is a constant which depends on the
density profile of the progenitor (\textit{S5}). The value of $\alpha$
is $\approx 10$ for a uniform density sphere or $\approx 30$ or
$\approx 90$ if the density profile of the sphere drops as $r^{-1}$ or
$r^{-2}$, respectively (\textit{S5}). If the system is better modeled
by a thin shell than a uniform sphere, then $\alpha=1$. We adopt
$\alpha \approx 10$ as we assume that the depth of the shock at the
time of the precursor is not negligible, and that the density profile
in the relevant part of the envelope is roughly constant. This seems
to be consistent with careful models of red-supergiant envelopes
(\textit{S6}).

Equating $\tau_{\rm diff}$ with $\tau_{\rm s}$ and rearranging
produces
\begin{equation}
d \approx \frac{\alpha c l}{3 v}
\end{equation}
in which we will substitute the mean free path for an opacity $\kappa$
dominated by electron scattering in a hydrogen atmosphere $\kappa_{\rm
es,H}$ of density $\rho$. This gives
\begin{equation}
d \approx 2.5 \times 10^{11}~{\rm{m}}~\left(\frac{\alpha}{10}\right)
{\left(\frac{\kappa}{\kappa_{\rm es,H}}\right)}^{-1}
{\left(\frac{\rho}{10^{-8}~{\rm kg~m^{-3}}}\right)}^{-1}
{\left(\frac{v}{10^{7}~{\rm m~s^{-1}}}\right)}^{-1}
\end{equation}
for the depth of the shock at the time when the radiative precursor
becomes visible at the surface. In the main text, we have omitted the
dependence on $\alpha$ and $\kappa$ for simplicity. Note that the
depth estimated here is a very good match to the depth of the shock in
Fig 3. at the start of the radiative precursor.

This value for $d$ is $\sim 350~R_{\odot}$. As the progenitor's
radius must be larger than $d$, we require a red supergiant, as
expected for a IIP SN. The duration of the radiative precursor should
be $d/v = 2.5 \times 10^{4}~{\rm s}$ for the parameters above,
i.e. almost 7 hours and thus in good agreement with our
observations. To be precise, we should increase this duration by the
light travel time across the disc, but this constitutes a fairly small
correction. As stated in the main text, the duration of shock breakout
from a blue supergiant is completely incompatible with our
observations (see also \textit{S7}).

Comparison with our numerical simulations and observations suggests
that a lower value of $\alpha$ might provide a more precise match to
the data, as the precursor itself lasts only $\approx 4$ hours before
shock breakout. A value of $\alpha \approx 5$ is a reasonable value,
intermediate between $\alpha = 1$ for a thin shell and $\alpha = 10$
for a sphere of uniform density.

The density adopted here is consistent with the envelopes of
red-supergiant models used in previous SN modeling, although they are
towards the lower end of expectations (\textit{S6, S8}). Note that the
modelling of red supergiant envelopes is uncertain. Convection becomes
inefficient near the surface of such stars, rendering mixing-length
theory inadequate. In addition, the boundary conditions of red giant
models should be carefully matched to models of the star's wind in
order to faithfully model the density profile of the envelope
(\textit{S6}).

\subsection*{S3.2 Estimated Energy Release}

We can estimate the total energy released in the radiative precursor
$E_{rp}$ in a way almost independent of the density profile. The total
energy radiated during the passage of a radiation-dominated shock with
velocity $v$ through a mass $M$ is $\approx (18/49) M v^{2}$ and we
can approximate the mass as $M \approx 4 \pi R^{2} \rho d$, where
$\rho$ and $d$ are again the density of the outer envelope and the
depth of the shock when the radiative precursor is first visible. We
take $R$ to be the radius of the star, which is a good approximation
if $R \gg d$. We can now replace $d$ using Eq. (S2) and then use $\rho
l = \kappa^{-1}$ to obtain
\begin{equation}
E_{\rm rp} \approx \left(\frac{18}{49}\right) 4 \pi R^{2} \frac{\alpha c
v}{3 \kappa} ,
\end{equation}
%
%%%%\left(\frac{\alpha}{10}\right)
\begin{equation}
E_{\rm rp} \approx 2.3 \times
10^{42}~{\rm{J}}~{\left(\frac{R}{10^{12}~{\rm m}}\right)}^{2}
{\left(\frac{\alpha}{10}\right)}
{\left(\frac{v}{10^{7}~{\rm m~s^{-1}}}\right)}^{-1}
{\left(\frac{\kappa}{\kappa_{\rm es,H}}\right)}^{-1} .
\end{equation}
Dividing this by a duration of $\sim 10^{4}~{\rm{s}}$ predicts a mean
luminosity of $\sim 10^{38}~{\rm{W}}$, consistent with our
extrapolation from the observed UV flux using a temperature of $\sim
10^{5}~{\rm K}$ (see section S4.2).

\section*{S4 Numerical Light Curve Models}

We have produced UV light curves for both the radiative precursor and
the post-shock-breakout adiabatic expansion. Our simple numerical
models use the physics essential to the respective phases. They
naturally produce light curves consistent with the data using the
expected physical input and the very minimum of parameter fitting.

The following models both produce a spectral energy distribution
(SED). The effects of cosmological redshift were applied to the SED
and the time axis of the expansion. The full {\it GALEX} filter
functions were used to define the NUV and FUV bands. The extinction of
the UV emission probably constitutes the biggest uncertainty in our
models. We have used measured values for the extinction (see section
S1) such that the NUV flux which reaches us is $\approx 1/3.8$ of the
emitted flux, and the FUV flux is $\approx 1/9.0$ of the
unextinguished value. However, even the `local' measurements of the
host galaxy's extinction are not guaranteed to be exactly the same as
those which would be appropriate for the SN (see section S1).

\subsection*{S4.1 The Radiative Precursor}

To model the radiative precursor we have written a bespoke
one-dimensional, two-temperature, hydrodynamic code. The code is
Eulerian; we have used 800 radial cells across the initial model
(hence with a typical cell size of $\sim 10^{9}~{\rm m}$) and 4000
cells in total. Radiation transport is handled during each timestep by
solving the diffusion equation for the internal energy $U_{\rm rad}$
of radiation within the moving radiation-dominated plasma, where the
the diffusion constant is $c/(3 \rho \kappa)$ and we assume that the
opacity $\kappa$ is mostly due to electron scattering inside a
hydrogen-dominated plasma. In addition to elastic Compton scattering,
the radiation treatment includes Compton cooling and bremsstrahlung
(\textit{S9}), and the hydrodynamics naturally incorporates advection
and adiabatic cooling. When a temperature is required for the emission
model we take the fourth root of the energy density of the radiation,
$T={(cU_{\rm rad}/4\sigma)}^{(1/4)}$.

The initial conditions specify a density distribution for the cold (10
eV) envelope and for the hot (1 MeV) core. The core properties were
chosen such as to eventually produce a shock moving with a
characteristic velocity of $1-2 \times 10^{7}~{\rm m~s^{-1}}$. Note
that this means that the first model in Fig.~3 is not at core
collapse, or directly taken from a stellar evolution code. The times
in Fig.~3 are approximately relative to shock breakout. 

This model includes the essential physics to describe the motion of
the shock through the star. However, modelling the exact spectral
energy distribution and luminosity of the radiative precursor would be
a much more complex task, partly as during shock breakout the
luminosity may be augmented by some non-thermal emission (\textit{S10}). To
the accuracy currently demanded by the data in the UV waveband, it
seems reasonable to approximate the emission as black-body. 

We do not attempt a full solution of the radiative-transfer problem,
but note that the photons which have diffused to the surface will
carry a temperature which was imprinted on them deeper in the star, as
the opacity is largely due to elastic scattering from electrons. We
estimate that the typical photon will have diffused from an optical
depth of $c/3v_{\rm shock}$ and therefore adopt this depth to define
the characteristic temperature for the black body. However, the
radiation flux at the surface will be somewhat lower than at the
optical depth where the photons originated (as the diffusion speed is
rather lower than the speed of light). Our models indicate that the
intensity of emission will be lower by up to a factor of five than
would be expected for this black-body temperature over the entire
surface. Given those uncertainties, the light curves in Fig.~3 have
been roughly rescaled down to show that expected shapes are consistent
with the data. The solid light curve in Fig.~3 has been multiplied by
$1/2.5$ and the dashed by $1/4$, both within the uncertainties of this
modelling.

In addition to any simplifications introduced in our modelling, we
note that most of the luminosity during the radiative precursor will
be emitted at higher energies than we directly observe; it is unclear
whether a significant fraction of those shorter-wavelength photons
will lead to the production of UV radiation through some indirect
route.

This model has the significant benefit of physical clarity, but more
detailed and complex work will be needed to fully exploit future
observations.

\subsection*{S4.2 After Shock-Breakout}

After the radiative precursor, {\it GALEX} has observed the early
stages of the SN's expansion. As summarized in the main text, this
phase is relatively simple to understand. The radiation-dominated
plasma expands freely (with almost constant velocity) and cools
adiabatically, hence $T \propto 1/R$ (\textit{S5}). The energy source
is the internal energy of the plasma; radioactivity is only relevant
much later.  Our model also assumes that the emission can be
approximated by a single-temperature black body, and this rest-frame
SED is converted into an observer-frame {\it GALEX} UV flux as
described at the start of section S4. The assumption that there is a
well-defined photosphere is reasonable for the very early stages of
expansion which we have observed.

The model light curves represented in Fig.~2 were produced by this
physical model of free adiabatic expansion. The initial radius and
expansion velocity were set by hand to a range of expected values for
the progenitor of such a type IIP SNe. Then the initial temperature
and time of explosion were fitted such that the $\chi^{2}$ parameter
with respect to the NUV data was minimized. The FUV data was not
fitted, but the model light curves are still consistent with the
data. This second phase, visible in the UV after the radiative
precursor, constrains the dimensions of the precursor independently of
the precursor model.

As the envelope cools and becomes less dense, the later behavior and
definition of the photosphere is more complex. In particular, the
plateau in the late-time optical light curve that is characteristic of
type IIP SNe is thought to be due to such complications. During the
plateau, the effective photosphere moves inwards in mass but remains
at an almost constant radial position and temperature. The color of
the SN is not precisely constant during the plateau (see Fig.~1); the
emission during that stage is not from a simple black body.

\subsection*{S4.3 The Second Peak in the UV Light Curve}

Although the physics governing the phase of adiabatic expansion is
simpler than in the previous or subsequent epochs, we find that the
main feature of this era is sometimes not intuitively
understood. Fig.~S4 demonstrates how the moving peak of the black-body
spectral energy distribution allows the luminosity in a particular
waveband to change non-monotonically, even though the total luminosity
is always decreasing.

As an alternative way to visualise this, we can write an equation for
the black body luminosity as a function of temperature and frequency,
$L(T, ~\nu)$, neglecting numerical factors, as
\begin{equation}
L(T,~\nu) \propto \nu^{3} R^{2} \frac{1}{e^{h \nu/kT} - 1},\
\end{equation}
where $h$ and $k$ are Plank's and Boltzmann's constants,
respectively. In the Rayleigh-Jeans limit this becomes
\begin{equation}
L(T,~\nu) \propto \nu^{2} T R^{2}  \propto \nu^{2} \frac{1}{T}
\end{equation}
where we have used the fact that $R \propto 1/T$ in the phase of
adiabatic expansion (\textit{S5}). So in the Rayleigh-Jeans portion of
the spectrum, the luminosity at a given frequency from the surface of
an adiabatically expanding optically-thick sphere is inversely
proportional to the temperature. The luminosity at that wavelength
begins to decline once the peak of the spectrum moves close to the UV
band and the Rayleigh-Jeans approximation is no longer valid at that
frequency.  Fig.~S4 suggests that the second peak occurs for a given
$\nu$ when $h \nu \approx 2kT$.

Note that this second peak in the light curve (the `adiabatic peak')
occurs at a different time for each frequency. The maximum bolometric
luminosity occurs at the only time when the luminosity in all
wavebands peaks simultaneously, and the secondary UV maximum is not
coincident with a peak in the visible light output. Furthermore, once
the phase of adiabatic expansion is over, our analysis is no longer
valid (e.g., the optical plateau is governed by completely different
physics).

There seems to be some confusion over whether the observations in
\textit{(S11)} have resolved shock breakout in the type Ib/c SN
1999ex. Note that the cadence of their observations is easily long
enough to miss the radiative precursor that we have observed. The
progenitor of 1999ex would not be a red supergiant but a much more
compact star. The duration of the radiative precursor preceding shock
emergence from such a compact star should be much shorter than that
which we observe. We thus find it extremely unlikely that
\textit{(S11)} observed the shock emerging from within the
star. However, we note that the timescale of the early dip they
observe in the $U$ band is consistent with the timescale we find for
the phase of adiabatic cooling.

\section*{S5 Suggestions for Further Observations}

Finding more events like SNLS-04D2dc will help further our
understanding of core-collapse SNe. We provide some general ideas on
how a larger sample of such events might be obtained. The starting
place for any such survey design must be the assumption that
SNLS-04D2dc was a normal event; we assume that a UV light curve of the
same absolute magnitude in the near-UV is associated with all -- or at
least a majority of -- Type II SNe. If this is the case, we must
further consider the dust extinction in the host galaxies and the
locations of Type II SNe. The near-UV is much more sensitive to dust
extinction than optical wavelengths, and so it is conceivable that a
SN Type II occurring in a heavily extincted host galaxy is detected in
optical filters, but remains undetected in the UV.

Taking our discovery of SNLS-04D2dc as the starting point, we can
estimate the rate of such events in a SN survey similar to
SNLS. SNLS-04D2dc is at the edge of what is detectable in single
\textit{GALEX} visits, so we can expect that no similar events will be
detectable beyond a redshift of $z \sim 0.2$ (the host galaxy of
SNLS-04D2dc is at $z = 0.1854$). We can compute the expected SN Type
II rate from the typical cosmic star formation density out to $z \sim
0.2$ probed by one \textit{GALEX} field of view (circular, 1.4 deg
diameter). The co-moving volume probed by this field of view is $\rm
\sim 8 \times 10^{4}~Mpc^{3}$. Combining this with a star formation
rate density of $\rho_{\rm SFR} \sim 0.03 ~{\rm
M_{\odot}yr^{-1}Mpc^{-3}}$ (\textit{S12}) yields a total star formation
rate probed of $\sim 2.5 \times 10^{3}~ {\rm
M_{\odot}yr^{-1}}$. Assuming a core-collapse rate of 1 per century per
$4 {\rm M_{\odot}yr^{-1}}$, this results in 6.2 events per year per
field of view. There are of course substantial uncertainties in this
estimate.

Thus, mounting a substantial \textit{GALEX} observational effort
covering e.g 4 fields-of-view continuously would yield about 2 such
events per month. A dedicated optical photometry and spectroscopic
survey for prompt follow-up of any detected events would also be
necessary. The resulting deep \textit{GALEX} image will also be
scientifically useful for other purposes.  The key feature of any
future similar observations must remain the high
cadence. \textit{GALEX} with its ultraviolet capability and 90-minute
orbit is the most suitable platform for further research into the
radiative precursors of supernovae.

\subsection*{References and Notes}
\begin{itemize}
\item[S1.]  Y.~C.~Pei, {\it ApJ}\/ {\bf 395} 130 (1992).
\item[S2.]
P.~Nugent, {\it et al., ApJ}\/ {\bf 645}, 841 (2006).
\item[S3.]
H.-J.~R\"oser, K.~Meisenheimer, {\it A\&A}\/ {\bf 252}, 458 (1991).
\item[S4.]
E.~Bertin, S.~Arnouts, {\it A\&AS}\/ {\bf 117}, 393 (1996).
\item[S5.]
D.~Arnett \textit{Supernovae and nucleosynthesis. An investigation of the history of matter, from the Big Bang to the present} Princeton University Press (1996)
\item[S6.]
A.~Heger, {\it PhD thesis}, Max-Planck-Institut f\"{u}r Astrophysik (1998).
\item[S7.]
C.~D.~Matzner, C.~F.~McKee, {\it ApJ}\/	{\bf 510}, 379 (1999). 
\item[S8.]
S.~W.~Falk, W.~D.~Arnett, {\it ApJS}\/ {\bf 33}, 515 (1977).
\item[S9.]
R.~A.~Chevalier, R.~I.~Klein, {\it ApJ}\/ {\bf 234}, 597, (1979).
\item[S10.]
R.~I.~Klein, R.~A.~Chevalier, {\it ApJ}\/ {\bf 223}, L109 (1978).
\item[S11.]
M.~Stritzinger, {\it et al., AJ}\/ {\bf 124}, 2100 (2002).
\item[S12.]
K.~ Glazebrook, {\it et al., ApJ}\/ {\bf 587}, 55  (2003). 
\end{itemize}

%%%%% SOM figures

\clearpage

\begin{center}
\includegraphics[angle=90, width=\textwidth]{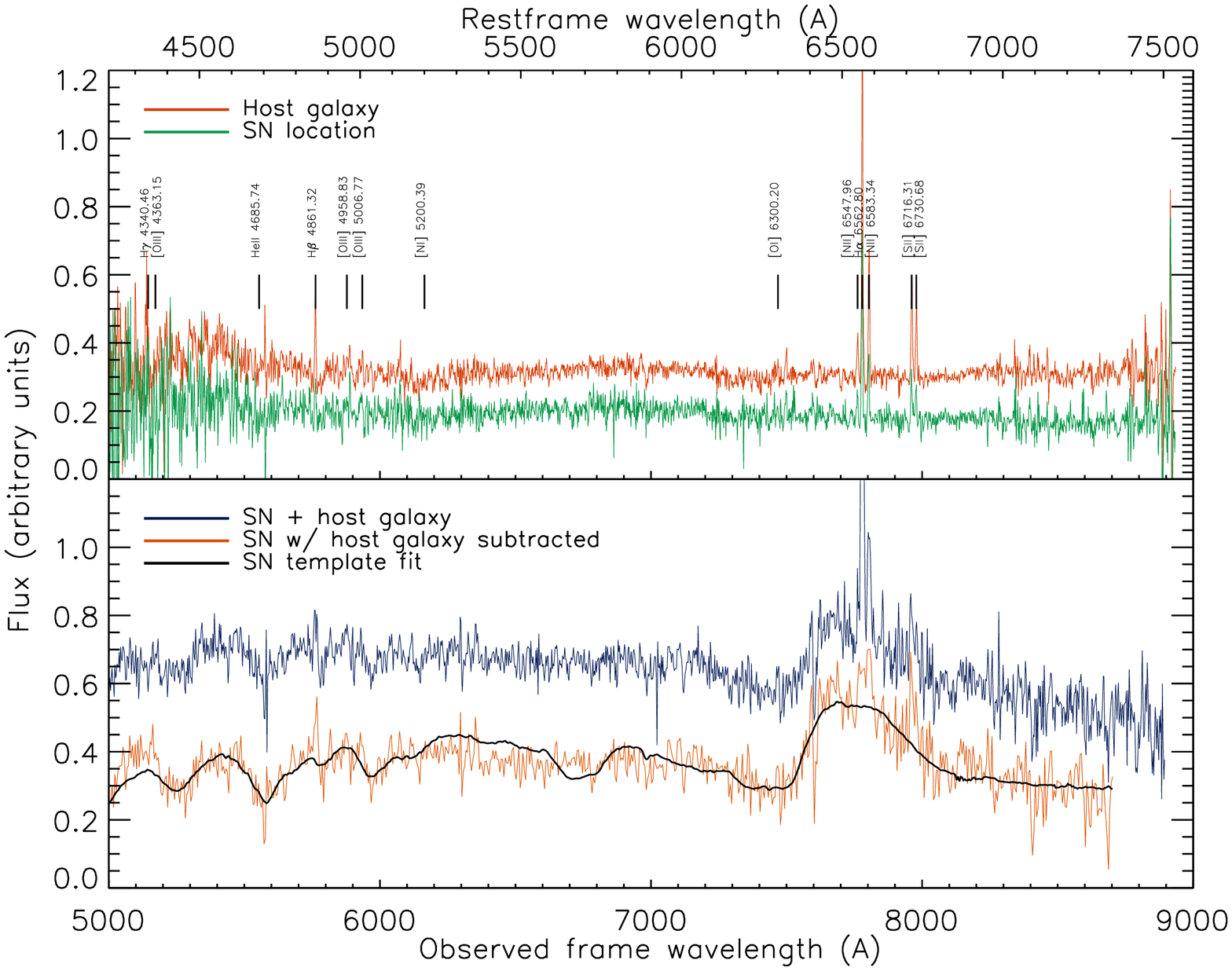}\\
\end{center}
\noindent {\bf Fig.~S1.} In the upper panel, we show the Gemini
spectra of the host galaxy as a whole and of the SN location. Both
show strong emission lines as expected of a spiral galaxy, including
H$\alpha$ and H$\beta$ from which we estimate the extinction affecting
the SN. In the lower panel we show the VLT spectrum of the SN, prior
to and after the subtraction of a host galaxy template, together with
an SN Type IIP template identifying the SN type.

\clearpage

\begin{center}
\includegraphics[angle=90, width=0.8\textwidth]{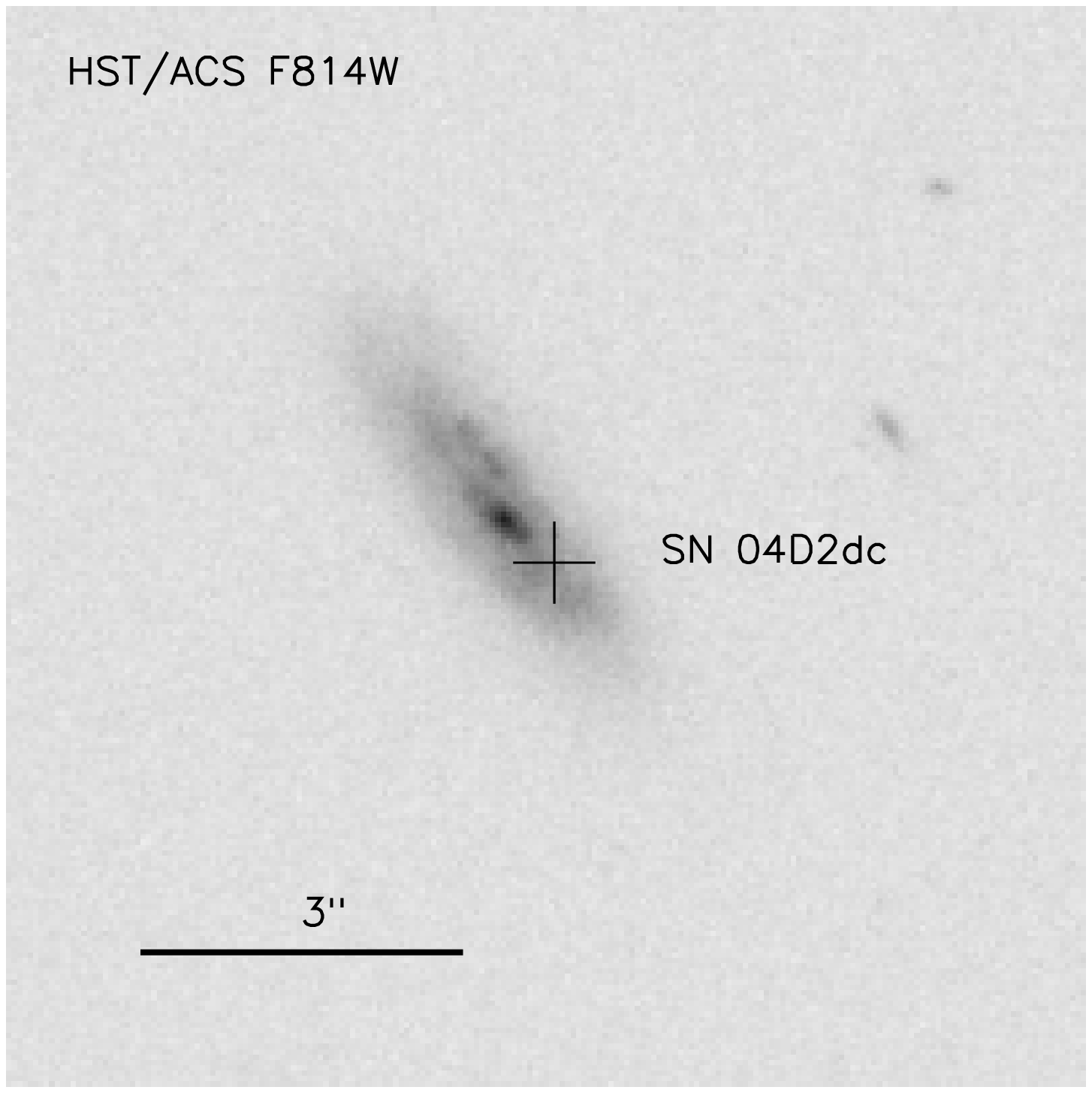}\\
\end{center}
\noindent {\bf Fig.~S2.} The \textit{Hubble Space Telescope}
F814W-band image of the host galaxy from the COSMOS survey. We
indicate the position of the SN and the image scale in arcseconds.

\clearpage

\begin{center}
\includegraphics[angle=270,width=0.8\textwidth]{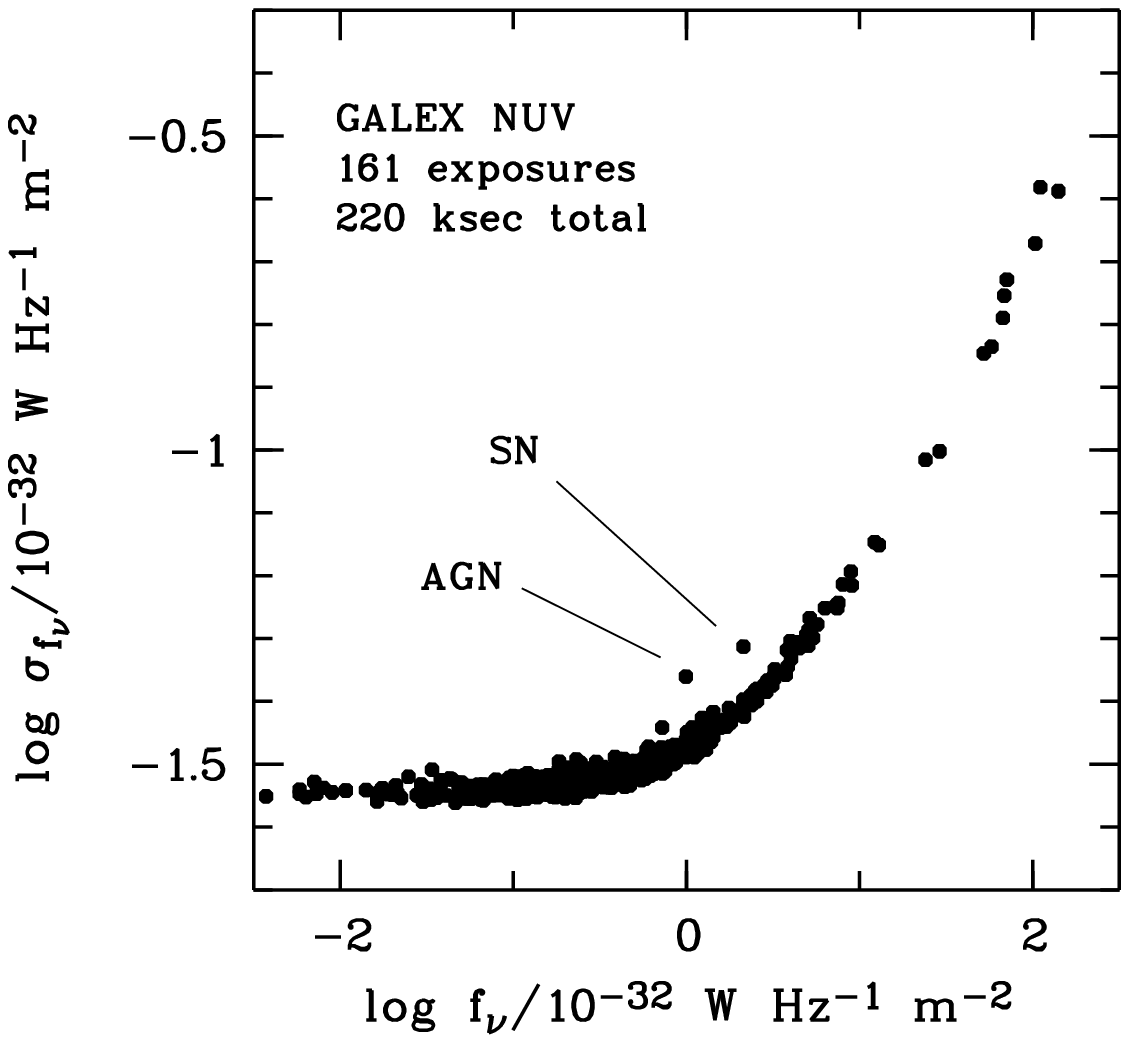}\\
\end{center}
\noindent {\bf Fig.~S3.}  Mean fluxes $f_\nu$ determined from 161
individual measurements and scatter $\sigma_{f_\nu}$ among them. Two
objects out of $\sim 1000$ are clearly variable as evidenced by their
increased scatter, one of which is the SN.

\clearpage

\begin{center}
\includegraphics[width=0.8\textwidth]{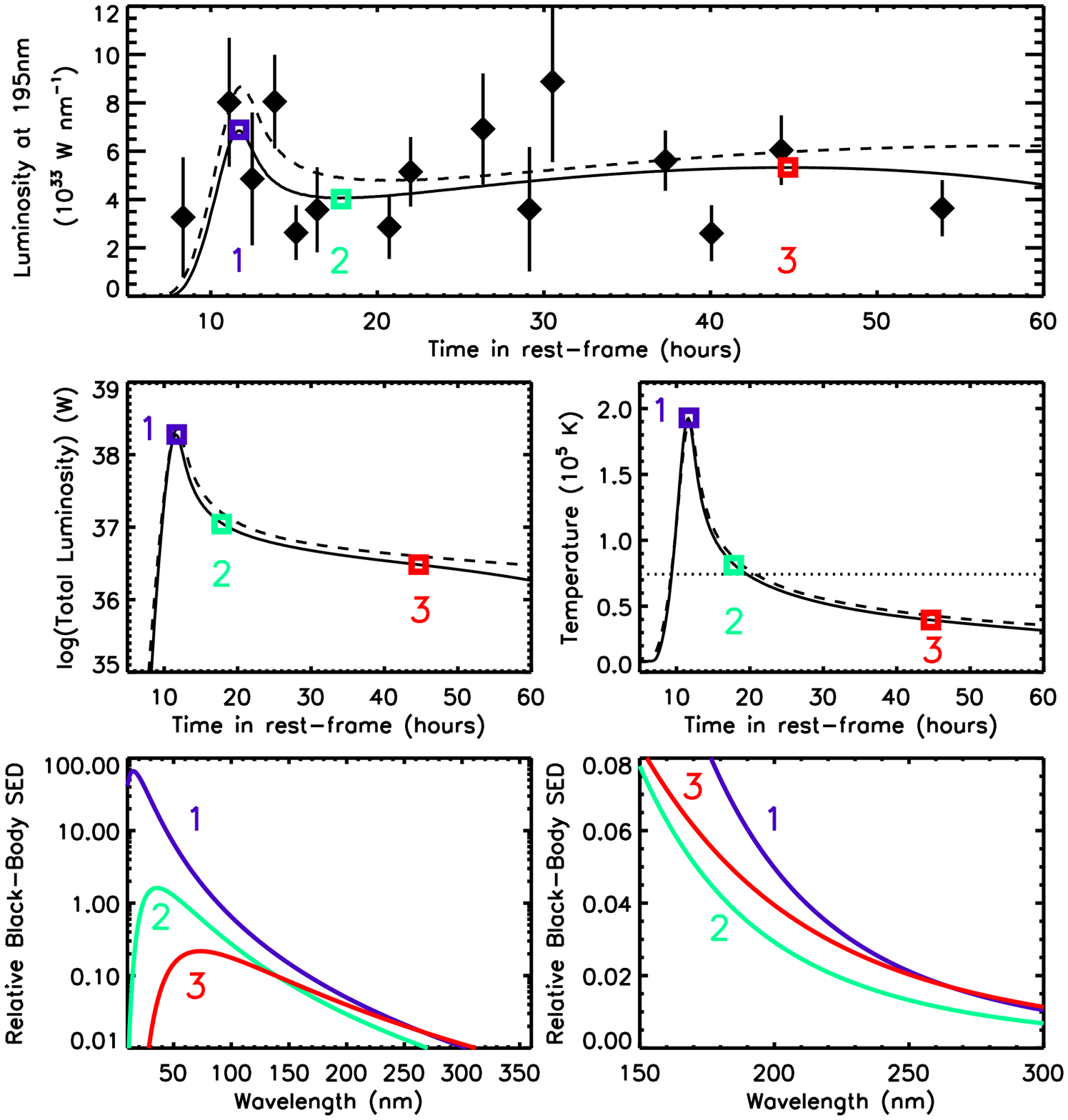}\\
\end{center}
\noindent {\bf Fig.~S4.} Illustration of the origin of the two peaks
in the UV light-curve for a monotonically decreasing bolometric
luminosity. The upper panel shows the same radiation-hydrodynamic
models as in Fig.~3 of the main text, again alongside the
\textit{GALEX} data. The model is shown going beyond the end of the
radiative precursor for which it was designed. However the model
continues to approximate the emission as coming from a black-body,
using the same assumptions adopted for the radiative precursor (as
described in S4.1). The three turning points for one of the model UV
light curves are marked with coloured squares and numbers. The middle
panels show the evolution of the total luminosity and emission
temperature, with the same three epochs marked with the same
colours. The horizontal dotted line represents $kT \approx 6.4{\rm
eV}$, i.e. it indicates the energy of a 195 nm photon. The lower two
panels show that, even though the overall area under the spectral
energy distribution (SED) decreases with time, the UV emission
displays non-monotonic behaviour. Between points one and two, the
shape of the light curve is dominated by the escape of the temperature
spike associated with the shock, and the UV luminosity drops along
with the total luminosity. Between the second and third points,
adiabatic expansion dominates, and the UV waveband is far enough from
the peak in the SED that the UV luminosity increases with decreasing
temperature (see also section S4.3). After the red point, the peak in
the SED is at a long enough wavelength that the Rayleigh-Jeans
approximation is no longer valid at 195 nm, and the luminosity at
195nm begins to fall with continued adiabatic expansion.  Eventually
the expansion will no longer be adiabatic, and the appearance of the
photosphere will be controlled by other processes, for example during
the optical plateau.
\clearpage

%%%%% SOM Tables

\begin{table}
\caption{GALEX Near-UV Photometry} \tiny
\label{tab:nuv1}
\begin{tabular}{@{}lrrrl}
\hline
\hline
MJD & NUV Flux    & NUV Flux Error  & \\
    & $10^{-32} \rm W m^2 Hz^{-1}$        & $10^{-32} \rm W m^2 Hz^{-1}$            &\\
\hline
 53045.47266  &      -0.188  &      0.4387&\\
 53045.53906  &      -0.407  &      0.4056&\\
 53045.60938  &       0.209  &      0.4793&\\
 53045.67969  &      -1.151  &      0.4016&\\
 53045.74609  &       0.350  &      0.4874&\\
 53045.81250  &      -0.028  &      0.4584&\\
 53046.42969  &       0.229  &      0.4769&\\
 53046.50000  &       0.038  &      0.4438&\\
 53046.57031  &      -0.270  &      0.4611&\\
 53046.63672  &       0.213  &      0.4619&\\
 53046.70312  &      -0.084  &      0.4672&\\
 53046.77344  &      -0.589  &      0.4133&\\
 53047.39062  &      -0.511  &      0.4381&\\
 53047.46094  &       0.313  &      0.4613&\\
 53047.52734  &       0.224  &      0.4848&\\
 53047.59375  &       0.621  &      0.4843&\\
 53047.66406  &      -0.085  &      0.4527&\\
 53048.34766  &      -0.800  &      0.4836&\\
 53048.41797  &       0.282  &      0.4888&\\
 53048.48438  &      -0.020  &      0.4807&\\
 53048.55469  &       0.266  &      0.4825&\\
 53048.62500  &       0.565  &      0.5023&\\
 53048.69141  &      -0.454  &      0.4489&\\
 53048.83594  &       0.210  &      0.6980&\\
 53049.23438  &       0.455  &      0.7840&\\
 53049.30469  &      -0.052  &      0.5565&\\
 53049.37500  &       0.450  &      0.5516&\\
 53049.44531  &      -0.324  &      0.4504&\\
 53049.51562  &      -0.554  &      0.4441&\\
 53049.65234  &      -0.250  &      0.4439&\\
 53049.71875  &       0.383  &      0.4735&\\
 53049.86328  &       0.960  &      0.8236&\\
 53050.33594  &      -0.176  &      0.5433&\\
 53050.40625  &       0.685  &      0.5213&\\
 53050.47266  &       0.365  &      0.4710&\\
 53050.54297  &      -0.111  &      0.4546&\\
 53050.60938  &      -0.332  &      0.4570&\\
 53050.67969  &       0.106  &      0.4796&\\
 53050.75000  &      -0.421  &      0.4401&\\
 53051.36328  &       0.574  &      0.5624&\\
 53051.43359  &       0.631  &      0.5159&\\
 53051.50000  &      -0.396  &      0.4509&\\
 53051.57031  &      -0.072  &      0.4402&\\
 53051.64062  &       0.222  &      0.4667&\\
 53051.70703  &       0.133  &      0.4544&\\
 53051.77734  &      -0.144  &      0.5503&\\
 53051.84766  &      -0.914  &      0.5761&\\
 53051.91797  &      -0.055  &      0.8137&\\
 53052.32031  &      -0.375  &      0.6040&\\
 53052.39062  &       0.022  &      0.5434&\\
 53052.46094  &      -0.397  &      0.4500&\\
 53052.53125  &       0.460  &      0.4855&\\
 53052.59766  &      -0.261  &      0.4578&\\
 53052.73438  &       0.098  &      0.4648&\\
 53052.80469  &      -0.288  &      0.5093&\\
 53052.87500  &      -0.071  &      0.7091&\\
 53053.34766  &      -0.146  &      0.5811&\\
 53053.41797  &       0.432  &      0.5636&\\
 53053.48828  &       0.094  &      0.4655&\\
 53053.55859  &      -0.177  &      0.4462&\\
 53053.62500  &       0.125  &      0.4438&\\
 53053.69531  &       0.248  &      0.4731&\\
 53053.83594  &      -0.005  &      0.5660&\\
 53053.97656  &       1.047  &      1.1173&\\
 53054.30469  &       0.058  &      0.6822&\\
 53054.44531  &      -0.280  &      0.4848&\\
 53054.51562  &      -0.226  &      0.4638&\\
 53054.58594  &      -0.156  &      0.4269&\\
 53054.65234  &      -0.195  &      0.4435&\\
 53054.72266  &      -0.140  &      0.4149&\\
 53054.79297  &       0.361  &      0.5057&\\
 53054.86328  &       0.100  &      0.5737&\\
\hline
\end{tabular}
\begin{flushleft}
\end{flushleft}
\end{table}

\begin{table*}
\caption{GALEX Near-UV Photometry (Continued.)}
\tiny
\label{tab:nuv2}
\begin{tabular}{@{}lrrrl}
\hline
\hline
MJD & NUV Flux    & NUV Flux Error  & \\
    & $10^{-32} \rm W m^2 Hz^{-1}$        & $10^{-32} \rm W m^2 Hz^{-1}$            &\\
\hline
 53055.33594  &       0.059  &      0.6200&\\
 53055.47656  &       0.203  &      0.4691&\\
 53055.54297  &       0.156  &      0.4366&\\
 53055.61328  &      -0.072  &      0.4289&\\
 53056.03125  &       0.417  &      0.8714&\\
 53056.50391  &       0.038  &      0.4269&\\
 53056.64062  &      -0.573  &      0.3995&\\
 53056.71094  &       0.171  &      0.4217&\\
 53056.77734  &      -0.221  &      0.4422&\\
 53056.84766  &      -0.107  &      0.5448&\\
 53056.91797  &       0.326  &      0.6477&\\
 53059.30859  &       1.422  &      0.9122&\\
 53059.37891  &      -0.986  &      0.5482&\\
 53059.51562  &      -0.797  &      0.3895&\\
 53059.58594  &      -0.048  &      0.3995&\\
 53059.65234  &      -0.853  &      0.3731&\\
 53059.72266  &      -0.048  &      0.4205&\\
 53059.79297  &       0.475  &      0.4902&\\
 53059.92969  &      -0.417  &      0.5031&\\
 53060.00000  &      -0.331  &      0.5835&\\
 53060.07422  &      -1.135  &      0.8677&\\
 53060.27734  &      -0.697  &      0.6799&\\
 53060.34375  &      -0.334  &      0.6917&\\
 53060.40625  &       0.438  &      0.4355&\\
 53060.47656  &       0.172  &      0.4383&\\
 53060.61328  &      -0.422  &      0.3738&\\
 53060.67969  &      -0.309  &      0.3964&\\
 53060.75000  &       0.306  &      0.4367&\\
 53060.89062  &       0.607  &      0.6784&\\
 53061.03125  &       0.050  &      0.6912&\\
 53061.23438  &       0.762  &      0.9179&\\
 53061.36719  &      -0.402  &      0.4040&\\
 53061.50391  &       0.485  &      0.4308&\\
 53062.05859  &       1.359  &      1.0288&\\
 53062.19531  &       3.335  &      1.1072&\\
 53062.26562  &       2.018  &      1.1415&\\
 53062.32812  &       3.346  &      0.8054&\\
 53062.39453  &       1.095  &      0.4696&\\
 53062.45703  &       1.486  &      0.7300&\\
 53062.67188  &       1.190  &      0.5463&\\
 53062.73438  &       2.139  &      0.5966&\\
 53062.94922  &       2.877  &      0.9542&\\
 53063.08594  &       1.496  &      1.0685&\\
 53063.15625  &       3.689  &      1.3829&\\
 53063.48828  &       2.332  &      0.5185&\\
 53063.62500  &       1.083  &      0.4808&\\
 53063.83203  &       2.511  &      0.5981&\\
 53064.31250  &       1.514  &      0.4822&\\
 53065.27344  &       1.609  &      0.5949&\\
 53065.33594  &       1.500  &      0.4635&\\
 53066.23047  &       1.415  &      0.5879&\\
 53066.29688  &       0.867  &      0.4614&\\
 53072.80469  &      -0.281  &      0.4055&\\
 53073.28125  &       0.735  &      0.4414&\\
 53073.96875  &       0.282  &      0.4158&\\
 53074.17188  &      -0.199  &      0.4207&\\
 53074.24219  &      -0.099  &      0.4227&\\
 53083.69531  &      -0.374  &      0.3748&\\
 53084.10156  &       0.120  &      0.5935&\\
 53084.58594  &      -0.112  &      0.4001&\\
 53084.65625  &       0.079  &      0.3968&\\
 53084.99609  &      -0.055  &      0.3947&\\
 53085.06250  &       0.289  &      0.6652&\\
 53085.33594  &      -0.023  &      0.5042&\\
 53085.40625  &      -0.122  &      0.4794&\\
 53085.88672  &      -0.327  &      0.3828&\\
 53763.30469  &      -0.606  &      0.5445&\\
 53763.78516  &       0.872  &      0.5973&\\
 53763.85938  &       2.163  &      0.8685&\\
 53764.25781  &       0.603  &      0.7651&\\
 53764.33203  &       1.417  &      0.6793&\\
 53764.40234  &       0.691  &      0.5247&\\
\hline
\end{tabular}
\begin{flushleft}
\end{flushleft}
\end{table*}

\begin{table*}
\caption{GALEX Near-UV Photometry (Continued)}
\tiny
\label{tab:nuv3}
\begin{tabular}{@{}lrrrl}
\hline
\hline
MJD & NUV Flux    & NUV Flux Error  & \\
    & $10^{-32} \rm W m^2 Hz^{-1}$        & $10^{-32} \rm W m^2 Hz^{-1}$            &\\
\hline
 53767.55469  &       0.539  &      0.4672&\\
 53769.81250  &      -0.384  &      0.4663&\\
 53769.95312  &       0.009  &      0.6820&\\
 53772.82812  &       0.145  &      0.4971&\\
 53775.98047  &      -0.459  &      0.4992&\\
 53776.11719  &      -0.128  &      0.6801&\\
 53783.30078  &       0.295  &      0.4269&\\
 53802.95312  &       0.223  &      0.4511&\\
 53812.60938  &      -1.282  &      0.3476&\\
 53813.77734  &      -0.097  &      0.4793&\\
 53813.84766  &      -0.132  &      0.6023&\\
 53814.17969  &      -1.397  &      0.8357&\\
 53814.25000  &      -0.970  &      0.5504&\\
 53814.32031  &       0.867  &      0.5290&\\
 53814.39062  &      -0.410  &      0.3981&\\
 53814.80469  &      -0.722  &      0.5056&\\
 53814.87500  &       0.502  &      0.7638&\\
\hline
\end{tabular}
\begin{flushleft}
\end{flushleft}
\end{table*}

\begin{table*}
\caption{SNLS Photometry}
\tiny
\label{tab:snls}
\begin{tabular}{@{}lrrrrl}
\hline
\hline
MJD & Filter &Flux    & Flux Error  & \\
    & &$10^{-32} \rm W m^2 Hz^{-1}$        & $10^{-32} \rm W m^2 Hz^{-1}$            &\\
\hline
   52993.597 & g &      -0.099  &      0.0416 &\\
   53026.613 & g &       0.001  &      0.0508 &\\
   53084.462 & g &       2.120  &      0.0574 &\\
   53105.378 & g &       1.051  &      0.0704 &\\
   53114.328 & g &       0.863  &      0.0828 &\\
   53118.369 & g &       0.789  &      0.1138 &\\
   53120.374 & g &       0.812  &      0.0558 &\\
   53136.332 & g &       0.602  &      0.0553 &\\
   53148.286 & g &       0.294  &      0.0662 &\\
   52993.576 & r &       0.064  &      0.0517 &\\
   53021.536 & r &       0.018  &      0.0546 &\\
   53025.458 & r &       0.034  &      0.0625 &\\
   53031.536 & r &       0.112  &      0.0888 &\\
   53081.362 & r &       2.951  &      0.0602 &\\
   53084.434 & r &       2.969  &      0.0767 &\\
   53093.418 & r &       2.546  &      0.1182 &\\
   53105.354 & r &       2.418  &      0.0900 &\\
   53114.296 & r &       2.366  &      0.0891 &\\
   53120.354 & r &       2.113  &      0.0666 &\\
   53136.312 & r &       1.922  &      0.0750 &\\
   53148.264 & r &       1.757  &      0.0861 &\\
   52993.537 & i &       0.061  &      0.0721 &\\
   53021.491 & i &       0.115  &      0.0675 &\\
   53026.472 & i &       0.092  &      0.0796 &\\
   53031.566 & i &       0.177  &      0.1400 &\\
   53080.445 & i &       3.007  &      0.1015 &\\
   53083.381 & i &       3.461  &      0.1081 &\\
   53093.369 & i &       3.405  &      0.1142 &\\
   53105.311 & i &       3.293  &      0.1024 &\\
   53114.267 & i &       3.264  &      0.1588 &\\
   53118.351 & i &       3.297  &      0.1126 &\\
   53135.254 & i &       2.654  &      0.2740 &\\
   53136.274 & i &       3.024  &      0.0718 &\\
   53147.329 & i &       2.798  &      0.0741 &\\
   52993.618 & z &       0.294  &      0.1470 &\\
   53094.413 & z &       2.961  &      0.1542 &\\
   53151.259 & z &       2.705  &      0.1786 &\\
\hline
\end{tabular}
\begin{flushleft}
\end{flushleft}
\end{table*}

\end{document}